\documentclass[aps,prc,showpacs,twocolumn,nofootinbib,groupedaddress,preprintnumbers]{revtex4-1}

\usepackage{calligra}
\usepackage{longtable}
\usepackage{graphicx}
\usepackage{dcolumn}
\usepackage{bm}
\usepackage{amsmath,amssymb}
\usepackage{mathrsfs}

\begin{document}

\preprint{YITP-15-71}


\title{Weak decay of $\Lambda_{c}^+$ for the study of $\Lambda(1405)$ and $\Lambda(1670)$}


\author{Kenta~Miyahara}
\email[]{miyahara@ruby.scphys.kyoto-u.ac.jp}
\affiliation{Department of Physics, Graduate School of Science, Kyoto University, Kyoto 606-8502, Japan}
\author{Tetsuo~Hyodo}
\affiliation{Yukawa Institute for Theoretical Physics, Kyoto University, Kyoto 606-8502, Japan}
\author{Eulogio~Oset}
\affiliation{Departamento de F\'{\i}sica Te\'orica and IFIC, Centro Mixto Universidad de Valencia-CSIC,
Institutos de Investigaci\'on de Paterna, Aptdo. 22085, 46071 Valencia,
Spain}


\date{\today}

\begin{abstract}                           

We study the $\Lambda_c$ decay process to $\pi^+$ and the meson-baryon final state for the analysis of $\Lambda$ resonances. Considering the Cabibbo-Kobayashi-Maskawa matrix, color suppression,  diquark correlation, and the kinematical condition, we show that the final meson-baryon state should be in a pure $I=0$ combination, when the meson-baryon invariant mass is small. Because the $I=1$ contamination usually makes it difficult to analyze $\Lambda$ resonances directly from experiments, the $\Lambda_{c}$ decay is an ideal process to study $\Lambda$ resonances. Calculating the final-state interaction by chiral unitary approaches, we find that the $\pi\Sigma$ invariant mass distributions have the same peak structure in the all charge combination of the $\pi\Sigma$ states related to the higher pole of the two poles of the $\Lambda(1405)$. Furthermore, we obtain a clear $\Lambda(1670)$ peak structure in the $\bar{K}N$ and $\eta\Lambda$ spectra.

\end{abstract}




\maketitle

\section{Introduction}  \label{sec:intro}  

The study of the $\Lambda(1405)$ resonance has been a hot topic in hadron physics, ever since it was predicted and observed more than 50 yr ago~\cite{Dalitz:1959dn,Dalitz:1960du,Alston:1961zzd}. There have been plenty of discussions to understand its mass, width, and the internal structure.

From a modern perspective, the $\Lambda(1405)$ is studied by the theoretical framework which combines the coupled-channel unitarity with chiral perturbation theory~\cite{Kaiser:1995eg,Oset:1997it,Oller:2000fj,Lutz:2001yb,Hyodo:2011ur}. A remarkable finding in this approach is the appearance of two poles in the complex energy plane near the $\Lambda(1405)$ resonance~\cite{Oller:2000fj,Jido:2003cb,Hyodo:2007jq}. This picture is confirmed by the systematic studies~\cite{Ikeda:2011pi,Ikeda:2012au,Guo:2012vv,Mai:2014xna} using the next-to-leading-order chiral interactions with the new experimental constraint by the SIDDHARTA Collaboration~\cite{Bazzi:2011zj,Bazzi:2012eq}, as well as by the comprehensive analysis of the $\pi\Sigma$ spectra in photoproduction~\cite{Roca:2013av,Roca:2013cca}. The origin of two poles is attributed to the two attractive components in the leading-order chiral interaction (Weinberg-Tomozawa term), the singlet and octet channels in the SU(3) basis~\cite{Jido:2003cb}, or the $\bar{K}N$ and $\pi\Sigma$ channels in the isospin basis~\cite{Hyodo:2007jq}. It is worth mentioning that there is another attraction in the strangeness $S=-1$ and the isospin $I=0$ sector, the SU(3) octet, or the $K\Xi$ channel. In fact, the $\Lambda(1670)$ resonance can also be generated in the same approach with a large coupling to the $K\Xi$ channel~\cite{Oset:2001cn}. The internal structure of the $\Lambda(1405)$ also draws much attention. The dominance of the $\bar{K}N$ molecular component in the $\Lambda(1405)$ has been shown from various aspects~\cite{Hyodo:2007np,Roca:2008kr,Hyodo:2008xr,Sekihara:2008qk,Sekihara:2010uz,Sekihara:2012xp,Miyahara:2015bya}. A recent lattice QCD study also supports the $\bar{K}N$ molecular structure of the $\Lambda(1405)$~\cite{Hall:2014uca}.

Because the resonance peak lies below the $\bar{K}N$ threshold, the $\Lambda(1405)$ decays exclusively into the $\pi\Sigma$ channel. Recently, detailed experimental studies of the $\Lambda(1405)$ in the $\pi\Sigma$ spectra have been performed in the low-energy production reactions, such as the photoproduction by the LEPS Collaboration~\cite{Niiyama:2008rt} and by the CLAS Collaboration~\cite{Moriya:2013eb,Moriya:2013hwg,Moriya:2014kpv}, and the proton-induced reaction by the HADES Collaboration~\cite{Agakishiev:2012xk}. In these studies, because the signal of the $\Lambda(1405)$ overlaps with the $\Sigma(1385)$ resonance, careful event selection is needed to eliminate the $\Sigma(1385)$ contribution. In addition, the isospin interference effect causes the difference of the $\pi^{+}\Sigma^{-}$ and $\pi^{-}\Sigma^{+}$ spectra~\cite{Nacher:1998mi},
which is, in fact, confirmed in the experimental data. This means that the charged $\pi\Sigma$ spectrum, whose detection is, in general, easier than the neutral one, is not in pure $I=0$. For the understanding of the property of the $\Lambda(1405)$, a detailed analysis is required to extract the neutral $\pi^{0}\Sigma^{0}$ channel, as done by the CLAS Collaboration.

Meanwhile, it has been shown in recent studies that the selection of the particular modes is possible in the nonleptonic weak decays of heavy hadrons~\cite{Chau:1982da,Chau:1987tk,Cheng:2010vk,Stone:2013eaa,Liang:2014tia,Bayar:2014qha,Xie:2014tma,Xie:2014gla,Liang:2014ama,Albaladejo:2015kea,Feijoo:2015cca}. Among others, of particular importance is the study of the $\Lambda_{b}\to J/\psi \Lambda(1405)$ decay, where the $\Lambda(1405)$ is generated in the final-state interaction of the meson-baryon pair~\cite{Roca:2015tea,Feijoo:2015cca}. It is found that the weak decay process near the $\Lambda(1405)$ production is dominated by the  $I=0$ meson-baryon pair. The $\Lambda_{b}\to J/\psi \Lambda(1405)$ process therefore provides a clear meson-baryon final-state interaction with almost no contamination from the  $\Sigma(1385)$ and the $I=1$ amplitude. This is different from the low-energy production experiments. Very recently, the $\Lambda_{b}\to J/\psi K^{-}p$ decay has been experimentally studied by the LHCb Collaboration~\cite{Aaij:2015tga} (see also Ref.~\cite{Roca:2015dva}) where the significant contribution from the $\Lambda^{*}$ resonances is observed in the $ K^{-}p$ spectrum.

In this paper, we propose to analyze the weak decay of the charmed baryon $\Lambda_{c}$ into $\pi^{+}MB$, where $MB$ stands for $\bar{K}N$, $\pi\Sigma$, or $\eta\Lambda$. This process has been discussed to extract the $\pi\Sigma$ scattering length from the cusp effect~\cite{Hyodo:2011js}. As we show below, the $\Lambda_{c}$ weak decay is also dominated by the $I=0$ $MB$ pair in the final state. In general, the production rate of $\Lambda_{c}$ should be much larger than that of $\Lambda_{b}$. In fact, the branching fractions of the $\Lambda_{c}\to \pi^{+}MB$ modes is experimentally well studied~\cite{Agashe:2014kda}. Furthermore, the decay of $\Lambda_c$ is more favored than the $\Lambda_b$ decay, as explained later.
It is also remarkable that the Belle Collaboration recently determined the absolute value of the branching fraction of the $\Lambda_{c}\to \pi^{+}K^{-}p$ mode~\cite{Zupanc:2013iki}. Moreover, the $K^{-}p$ spectrum in the $\pi^{+}K^{-}p$ decay has already been measured experimentally~\cite{Aitala:1999uq}. These facts indicate that the $\Lambda_{c}\to\pi^{+}MB$ process is an ideal channel by which to study the $\Lambda(1405)$. The dominance of the $I=0$ contribution is also advantageous for the study of the $\Lambda(1670)$ to suppress the unwanted $\Sigma^{*}$ contributions.

This paper is organized as follows. In Sec.~\ref{sec:formulation}, we present the theoretical formulation of the decay process $\Lambda_{c}\to\pi^{+}MB$. The formula for the invariant mass distribution will be given. The numerical results of the spectra are shown in Sec.~\ref{sec:result}, focusing on the $\Lambda(1405)$ and the $\Lambda(1670)$ separately. A discussion on the decay branching fractions of $\Lambda_{c}$ is also presented. A summary of this work is given in the last section.

\section{Formulation}  \label{sec:formulation}  

\begin{figure}[tb]
\begin{center}
\includegraphics[width=8cm,bb=0 0 642 379]{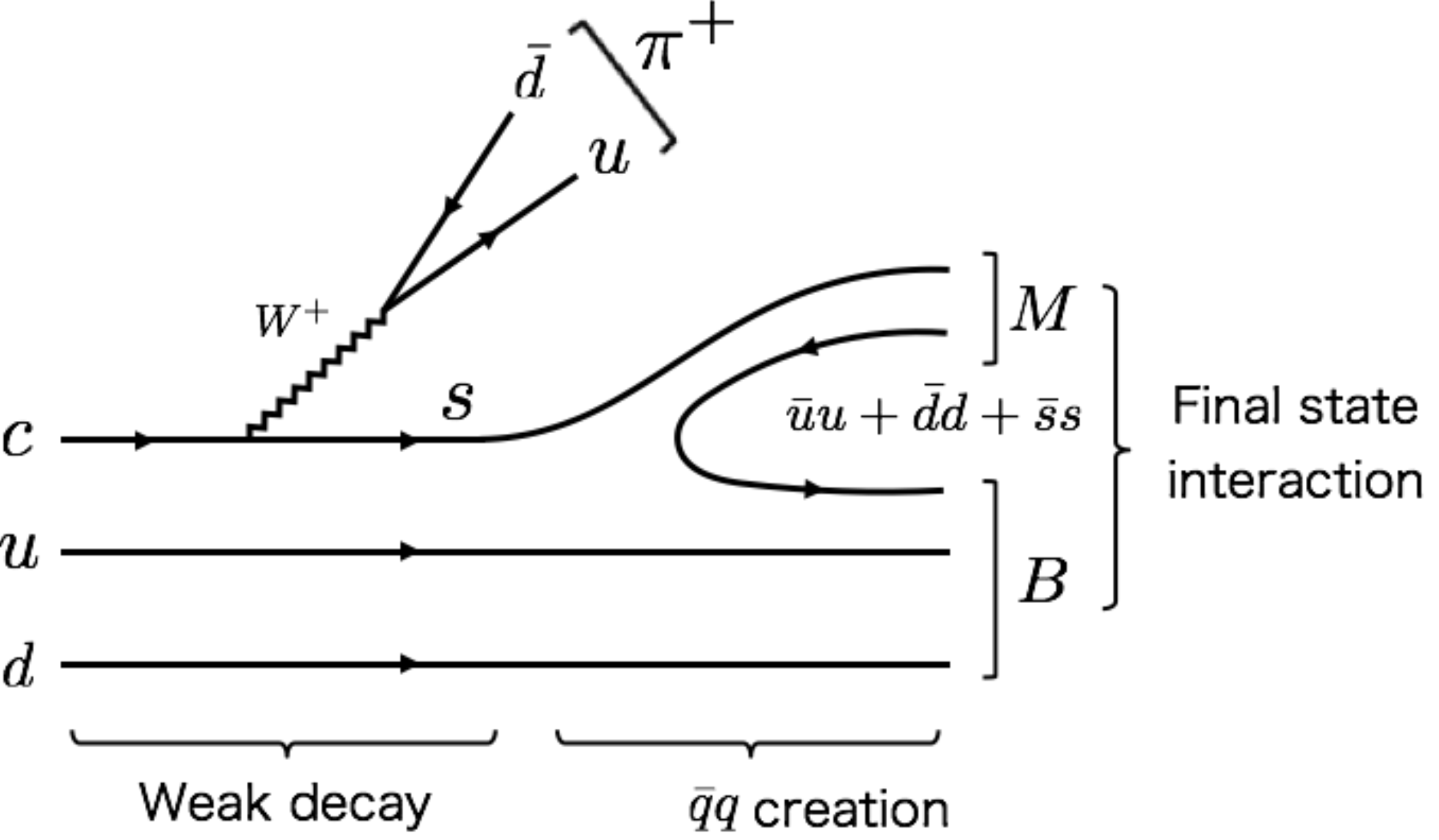}
\caption{The dominant diagram for the $\Lambda_c^+\to \pi^+MB$ decay. The solid lines and the wiggly line show the quarks and the $W$ boson, respectively. }
\label{fig:Lambdac_decay}  
\end{center}
\end{figure}

We consider the decay process $\Lambda_c^+\to\pi^+\Lambda^*\to\pi^+MB$, where $MB$ stands for the final meson-baryon states such as $\pi\Sigma$ and $\bar{K}N$. We show that, when the $MB$ invariant mass is restricted in the $\Lambda(1405)$ region, the dominant contribution of this decay process is given by the diagram shown in Fig.~\ref{fig:Lambdac_decay}. 
First, the charm quark in $\Lambda_c^+$ turns into the strange quark with the $\pi^+$ emission by the weak decay. Second, the $\bar{q}q$ creation occurs to form $M$ ($B$) from the $s$ quark ($ud$ diquark). Finally, considering the final-state interactions of the hadrons, we obtain the invariant mass distribution for the $\Lambda_c^+\to\pi^+MB$ process. In the following, we discuss these three steps separately.

\subsection{Weak decay}

\begin{figure}[tb]
\begin{center}
\includegraphics[width=8cm,bb=0 0 1066 228]{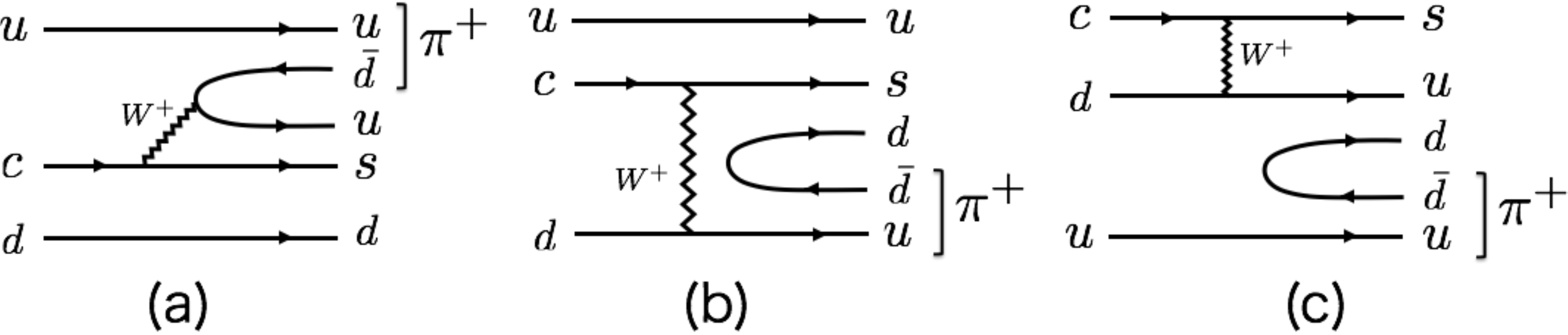}
\caption{The weak decay part of diagrams for the $\Lambda_c^+\to \pi^+MB$ decay except for Fig.~\ref{fig:Lambdac_decay}. The solid lines and the wiggly line show the quarks and the $W$ boson, respectively.}
\label{fig:other_decay}  
\end{center}
\end{figure}

We first discuss the decay of $\Lambda_{c}$ to produce $\pi^{+}$ and the $sud$ cluster in the final state. The Cabibbo favored weak processes are given by
\begin{align}
   &c\to s + u+\bar{d} \quad \text{: $c$ decay}\label{eq:cdecay} ,\\
   &c + d \to s + u \quad \text{: $cd$ scattering}\label{eq:cdscatt} .
\end{align}
The diagram shown in Fig.~\ref{fig:Lambdac_decay} is obtained by the $c$  decay process. Another contribution with the $c$ decay is to form $\pi^{+}$ using the $u$ quark in $\Lambda_{c}$ [Fig.~\ref{fig:other_decay}(a)]. In this process, however, the color of the $u\bar{d}$ pair from the $W^{+}$ decay is constrained to form the color singlet $\pi^{+}$. This process is therefore suppressed by the color factor in comparison with Fig.~\ref{fig:Lambdac_decay}. In addition, because the $ud$ diquark in $\Lambda_{c}$ is the most attractive ``good'' diquark~\cite{Jaffe:2004ph}, the process to destruct the strong diquark correlation [Fig.~\ref{fig:other_decay}(a)] is not favored. The contribution from the $cd$ scattering [Eq.~\eqref{eq:cdscatt}] [Figs.~\ref{fig:other_decay}(b) and \ref{fig:other_decay}(c)] is suppressed by the additional creation of a $\bar{q}q$ pair not attached to the $W$ boson as well as the $1/N_{c}$ suppression, compared with Fig.~\ref{fig:Lambdac_decay}. Diagrams ~\ref{fig:other_decay}(b) and ~\ref{fig:other_decay}(c) are called ``absorption diagrams" in the classification of Ref.~\cite{Chau:1982da}, and they are two-body processes, involving two quarks of the original $\Lambda_c$, 
which are suppressed compared to the one-body process (Fig.~\ref{fig:Lambdac_decay}) involving only the $c$ quark, the $u$, $d$ quarks acting as 
spectators. A discussion of this suppression is given in Ref.~\cite{Xie:2014tma}.

As discussed in Ref.~\cite{Hyodo:2011js}, the kinematical condition also favors the diagram shown in Fig.~\ref{fig:Lambdac_decay}. When the $\Lambda_{c}$ decays into the $\pi^{+}$ and $MB$ system with the invariant mass of 1400 MeV, the three-momentum of the final state is $\sim 700$ MeV in the rest frame of $\Lambda_{c}$. Thus, the $\pi^{+}$ should be emitted with a large momentum. It is kinematically favored to create the fast pion from the quarks involved by the weak process, because of the large mass of the $c$ quark. The diagrams Fig.~\ref{fig:other_decay}(a) and \ref{fig:other_decay}(c) are suppressed because one of the spectator quarks is required to participate in the $\pi^{+}$ formation.

In this way, the diagram in Fig.~\ref{fig:Lambdac_decay} is favored from the viewpoint of the Cabibbo-Kobayashi-Maskawa (CKM) matrix, color suppression, the diquark correlation, and the kinematical condition. We note that this diagram has a bigger strength than the dominant one of the $\Lambda_{b}\to J/\psi \Lambda(1405)$ decay \cite{Roca:2015tea}, in which the weak process contains the necessary Cabibbo suppressed $b\to c$ transition and proceeds via internal emission \cite{Chau:1982da}, where the color of every quark in the weak process is fixed.

In this process, because the $ud$ diquark in $\Lambda_{c}$ is the spectator, the $sud$ cluster in the final state is combined as 
\begin{align}
\frac{1}{\sqrt{2}}|s(ud-du)\rangle .  \notag
\end{align}
This combination is a pure $I=0$ state. Because the $\bar{q}q$ creation does not change the isospin, we conclude that the dominant contribution for the $\Lambda_{c}\to \pi^{+}MB$ process produces the $MB$ pair in $I=0$. We note that the unfavored diagrams that we neglect can produce the $sud$ state with $I=1$. We revisit the $I=1$ contribution at the end of Sec.~\ref{sec:result}.

\subsection{$\bar{q}q$ creation}

To create the $MB$ final state, we must proceed to hadronize the $sud$ state, creating an extra $\bar{q}q$ pair. Because the total spin parity of the $MB$ pair is $J^P=1/2^-$, the $s$ quark should be in $L=1$ after the $c$-quark decay, together with the spectator $ud$ diquark. To achieve the final $MB$ state where all quarks are in the ground state, the $\bar{q}q$ creation must involve the $s$ quark to deexcite into $L=0$. Then the hadronization proceeds as depicted in Fig.~\ref{fig:Lambdac_decay}, where the $s$ quark goes into the meson $M$ and the $ud$ diquark is used to form the baryon $B$. Another possibility, the formation of the baryon from the $s$ quark, is not favored because of the correlation of the good $ud$ diquark and the suppression discussed above by forcing a spectator quark from the $\Lambda_c$ to form the emerging meson.


To evaluate the relative fractions of the $MB$ state, we follow the same procedure as in Ref.~\cite{Roca:2015tea}. The flavor wavefunction of the final meson-baryon state $|MB\rangle$ is given by 
\begin{align}
|MB\rangle &= \frac{1}{\sqrt{2}}|s(\bar{u}u+\bar{d}d+\bar{s}s)(ud-du)\rangle  \notag \\
&= \frac{1}{\sqrt{2}}\sum_{i=1}^3|P_{3i}q_i(ud-du)\rangle,  \notag
\end{align}
where $q$ and $P$ represent the quark field and the $q\bar{q}$ pair, 
\begin{align}
q&\equiv
\left( \begin{array}{c}
u \\ d \\ s
\end{array} \right), \notag \\
P&\equiv q\bar{q}=
\left( \begin{array}{ccc}
u\bar{u}  &  u\bar{d}  &  u\bar{s}  \\
d\bar{u}  &  d\bar{d}  &  d\bar{s}  \\
s\bar{u}  &  s\bar{d}  &  s\bar{s}  
\end{array} \right).  \notag
\end{align}
Using the mesonic degrees of freedom, $P$ can be written as
\begin{align}
P=\left( \begin{array}{ccc}
\frac{\pi^0}{\sqrt{2}}+\frac{\eta}{\sqrt{3}}+\frac{\eta^\prime}{\sqrt{6}}  &  \pi^+  &  K^+  \\
\pi^-  &  -\frac{\pi^0}{\sqrt{2}}+\frac{\eta}{\sqrt{3}}+\frac{\eta^\prime}{\sqrt{6}}  &  K^0  \\
K^-  &  \bar{K}^0  &  -\frac{\eta}{\sqrt{3}}+\frac{2\eta^\prime}{\sqrt{6}}
\end{array} \right),  \notag
\end{align}
where we assume the ordinary mixing between the SU(3) singlet and octet states for $\eta$ and $\eta^\prime$~\cite{Bramon:1992kr}. Considering the overlap with the mixed antisymmetric flavor states of the baryons, we relate the three quark states with baryons as
\begin{align}
|p\rangle &=\frac{1}{\sqrt{2}}|u(ud-du)\rangle, \notag \\
|n\rangle &=\frac{1}{\sqrt{2}}|d(ud-du)\rangle, \notag \\
|\Lambda\rangle &=\frac{1}{\sqrt{12}}|(usd-dsu)+(dus-uds)+2(sud-sdu)\rangle. \notag
\end{align}
Using these hadronic representations, we obtain the meson-baryon states after the $\bar{q}q$ pair production as 
\begin{align}
|MB\rangle &= |K^-p\rangle + |\bar{K}^0n\rangle -\frac{\sqrt{2}}{3}|\eta\Lambda\rangle.  \label{eq:hadronstate} 
\end{align}
Here we neglect the irrelevant $\eta^{\prime}\Lambda$ channel because its threshold is above 2 GeV. We can see that we obtain the isospin $I=0$ $\bar{K}N$ combination in the phase convention that we use where $|K^-\rangle = -|I=1/2,I_z=-1/2\rangle$. 

\begin{figure}[tb]
\begin{center}
\includegraphics[width=8cm,bb=0 0 744 626]{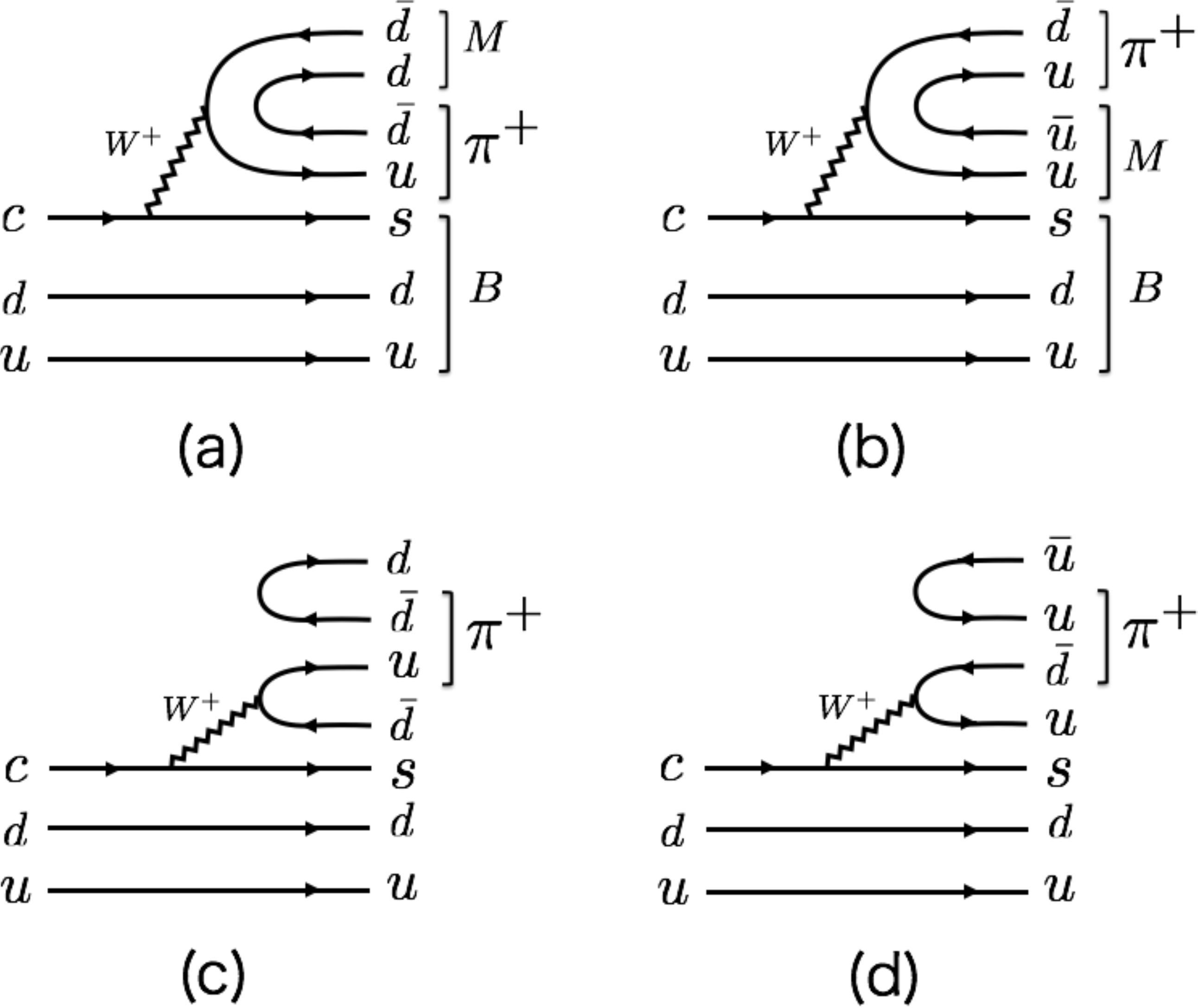}
\caption{Other diagrams for the $\Lambda_c^+\to \pi^+MB$ decay, which do not produce $\pi^+$ and $sud$ cluster. The solid lines and the wiggly line show the quarks and the $W$ boson, respectively.}
\label{fig:other_decay2}  
\end{center}
\end{figure}

Aside from the decay to $\pi^+$ and the $sud$ cluster, there are some candidates for the $\Lambda_c\to \pi^+MB$ decay process, as shown in Fig.~\ref{fig:other_decay2}. In the diagrams Figs.~\ref{fig:other_decay2}(c) and \ref{fig:other_decay2}(d), a quark of the $u\bar{d}$ pair from the weak decay should couple to the quarks of $sud$ cluster. Therefore, these diagrams are suppressed by the color factor.
From the diagrams Figs.~\ref{fig:other_decay2}(a) and \ref{fig:other_decay2}(b), only the $\eta\Lambda$ state appears as the final $MB$ pair. This is because the baryon is fixed as $B=\Lambda$, and the flavor wavefunction of the $M\pi^+$ state is obtained as follows:  
\begin{align}
|M\pi^+\rangle = |u(\bar{u}u+\bar{d}d)\bar{d}\rangle\propto |\eta\pi^++\eta^\prime\pi^+\rangle. \notag
\end{align}
The $\eta\Lambda$ state does not change the $I=0$ filter discussed in the previous section. In the numerical calculation, we find that the quantitative results do not strongly depend on the strength of the $\eta\Lambda$ fraction [this would change the coefficient of $\eta \Lambda$ in Eq.~\eqref{eq:hadronstate}, which does not play a very important role]. Hence, we do not consider the effect of these diagrams.

\subsection{Final-state interaction}

\begin{figure}[tb]
\begin{center}
\includegraphics[width=8cm,bb=0 0 911 169]{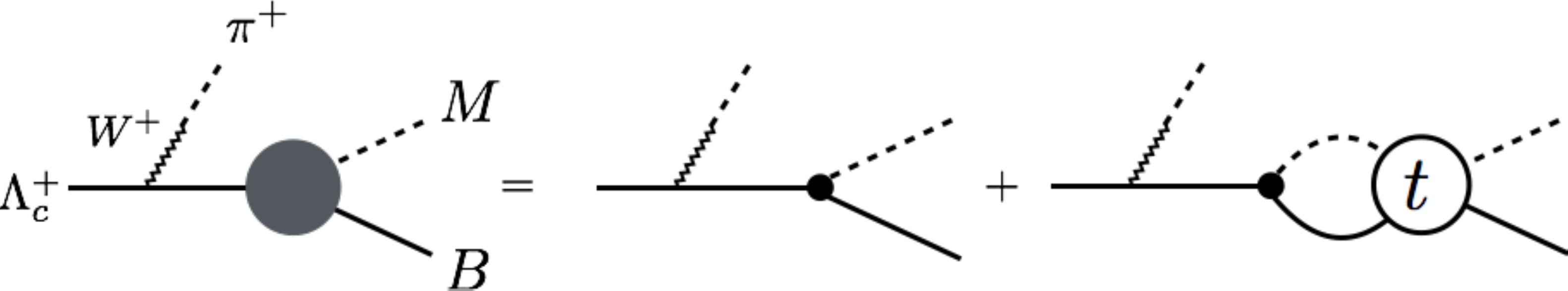}
\caption{The diagram for the meson-baryon final-state interaction (solid circle) as the sum of the tree part (dot) and the rescattering part with the meson-baryon scattering amplitude (open circle).}
\label{fig:FSI}  
\end{center}
\end{figure}

Here we derive the decay amplitude $\mathscr{M}$, taking the final-state interaction of the $MB$ pair into account. As shown in Fig.~\ref{fig:FSI}, the final-state interaction consists of the tree part and the rescattering part. 
The rescattering of the $MB$ pair is described by the chiral unitary approach~\cite{Kaiser:1995eg,Oset:1997it,Oller:2000fj,Lutz:2001yb,Hyodo:2011ur}, which is based on the chiral Lagrangians and is constructed to treat the nonperturbative phenomena. Though only the $K^-p,\ \bar{K^0}n,\ \eta\Lambda$ states appear in Eq.~\eqref{eq:hadronstate} in the tree-level production, the coupled-channel scattering leads to the other $MB$ states, $\pi^0\Sigma^0$, $\pi^-\Sigma^+$, $\pi^+\Sigma^-$, $\pi^{0}\Lambda$, $K^-p$, $\bar{K^0}n$, $\eta\Lambda$, $\eta\Sigma^0$, $K^+\Xi^-$, $K^0\Xi^0$.\footnote{The $\pi^{0}\Lambda$ and $\eta\Sigma^0$ channels are accessible only through the isospin breaking processes.} The decay amplitude for the $\Lambda_{c}\to \pi^{+}(MB)_{j}$ with the meson-baryon channel $j$ can then be written as
\begin{align}
\mathscr{M}_j =V_P &\left( h_j + \sum_i h_i G_i(M_{\rm inv})t_{ij}(M_{\rm inv}) \right),  \label{eq:amplitude} \\
h_{\pi^0\Sigma^0}&=h_{\pi^-\Sigma^+}=h_{\pi^+\Sigma^-}=h_{\pi^0\Lambda}=0,  \notag \\
h_{K^-p}&=h_{\bar{K^0}n}=1,  \notag \\
h_{\eta\Lambda}&=-\frac{\sqrt{2}}{3},  \notag  \\
h_{\eta\Sigma^0}&=h_{K^+\Xi^-}=h_{K^0\Xi^0}=0,  \notag
\end{align}
where $M_{\rm inv}$ represents the invariant mass of the meson-baryon states. The weak decay and the $\bar{q}q$ pair creation are represented by the factor $V_P$, which is assumed to be independent of the invariant mass $M_{\rm inv}$ in the limited range of invariant masses that we consider. The coefficients $h_j$ are derived from Eq.~\eqref{eq:hadronstate}. $G_j$ and $t_{ij}$ are respectively the meson-baryon loop function and scattering matrix calculated from chiral unitary approach. Explicit forms can be found in Refs.~\cite{Kaiser:1995eg,Oset:1997it,Oller:2000fj,Lutz:2001yb,Hyodo:2011ur}. In Eq.~\eqref{eq:amplitude}, the first term $h_j$ stands for the primary production (tree level diagram) and the second term accounts for the rescattering of the $MB$ states into the final-state channel $j$. It is also instructive for practical calculations to show the amplitude in the isospin basis. If we assume the isospin symmetry, the amplitudes of the decay to the $\pi\Sigma$ and $\bar{K}N$ channels are written, respectively, as 
\begin{align}
\mathscr{M}_{\pi^0\Sigma^0}&=\mathscr{M}_{\pi^-\Sigma^+}=\mathscr{M}_{\pi^+\Sigma^-}  \notag \\
=&V_P\left(-\sqrt{\frac{2}{3}} G_{\bar{K}N}t^{I=0}_{\bar{K}N,\pi\Sigma}+\frac{\sqrt{2}}{3\sqrt{3}}G_{\eta\Lambda}t^{I=0}_{\eta\Lambda,\pi\Sigma} \right),  \label{eq:amplitude_isobasis_piSig} \\
\mathscr{M}_{K^-p}&=\mathscr{M}_{\bar{K^0}n}  \notag  \\
=&V_P\left(1+G_{\bar{K}N}t^{I=0}_{\bar{K}N,\bar{K}N}-\frac{1}{3}G_{\eta\Lambda}t^{I=0}_{\eta\Lambda,\bar{K}N} \right).  \label{eq:amplitude_isobasis_KbarN}
\end{align}

The partial decay width of the $\Lambda_{c}$ into the $\pi^{+}(MB)_{j}$ channel is given by
\begin{align}
\Gamma_j
&=\int d\Pi_{3}|\mathscr{M}_j|^2 ,
\end{align}
where $d\Pi_{3}$ is the three-body phase space. The invariant mass distribution is obtained as the derivative of the partial width with respect to $M_{\rm inv}$. In the present case, because the amplitude $\mathscr{M}_j$ depends only on $M_{\rm inv}$, the mass distribution ${\rm d}\Gamma_{j}/{\rm d}M_{\rm inv}$ is obtained by integrating the phase space as
\begin{align}
\frac{{\rm d}\Gamma_j}{{\rm d}M_{\rm inv}} &=\frac{1}{(2\pi)^3}\frac{p_{\pi^+}\tilde{p}_jM_{\Lambda_c^+}M_j}{M_{\Lambda_c^+}^2} |\mathscr{M}_j|^2,  \label{eq:massdis}
\end{align}
where $M_j$ is the baryon mass. $p_{\pi^+}$ and $\tilde{p}_j$ represent the magnitude of the three-momentum of the emitted $\pi^+$ by the weak decay in the $\Lambda_c$ rest frame and of the meson of the final meson-baryon state in the meson-baryon rest frame,
\begin{align}
p_{\pi^+}=& \frac{\lambda^{1/2}(M_{\Lambda_c^+}^2,m_{\pi^+}^2,M_{\rm inv}^2)}{2M_{\Lambda_c^+}},\ \tilde{p}_j= \frac{\lambda^{1/2}(M_{\rm inv}^2,M_j^2,m_j^2)}{2M_{\rm inv}},  \notag \\
&\lambda(x,y,z) = x^{2}+y^{2}+z^{2}-2xy-2yz-2zx,  \notag
\end{align}
where $m_j$ represents the meson mass. 

Because the $\Lambda(1405)$ is mainly coupled to the $\pi\Sigma$ and $\bar{K}N$ channels, we calculate the invariant mass distribution of the decay to the  $\pi\Sigma$ and $\bar{K}N$ channels. For the study of the $\Lambda(1670)$, we also calculate the decay to the $\eta \Lambda$ channel.

\section{Results}  \label{sec:result}  

We present the numerical results of the $MB$ invariant mass spectra in the $\Lambda_{c}\to \pi^{+}MB$ decay. We first show the results in the energy region near the $\bar{K}N$ threshold where the $\Lambda(1405)$ resonance plays an important role. We then discuss the spectra in the higher energy region with the emphasis of the $\Lambda(1670)$ resonance. The decay branching fractions of $\Lambda_{c}$ into different final states are discussed at the end of this section.

\subsection{Spectrum near the $\bar{K}N$ threshold}

To calculate the region near the $\bar{K}N$ threshold quantitatively, the final-state interaction of the $MB$ system should be constrained by the new experimental data from the SIDDHARTA Collaboration~\cite{Bazzi:2011zj,Bazzi:2012eq}, because the precise measurement of the energy-level shift of kaonic hydrogen significantly reduces the uncertainty of the scattering amplitude below the $\bar{K}N$ threshold. Here we employ the meson-baryon amplitude in Refs.~\cite{Ikeda:2011pi,Ikeda:2012au}, which implements the next-to-leading-order terms in chiral perturbation theory to reproduce the low-energy $\bar{K}N$ scattering data, including the SIDDHARTA constraint. The isospin symmetry breaking is introduced by the physical values for the hadron masses. In this model, the two resonance poles of the $\Lambda(1405)$ are obtained at $1424-26i$ MeV and $1381-81i$ MeV.

\begin{figure}[tb]
\begin{center}
\includegraphics[width=8cm,bb=0 0 846 594]{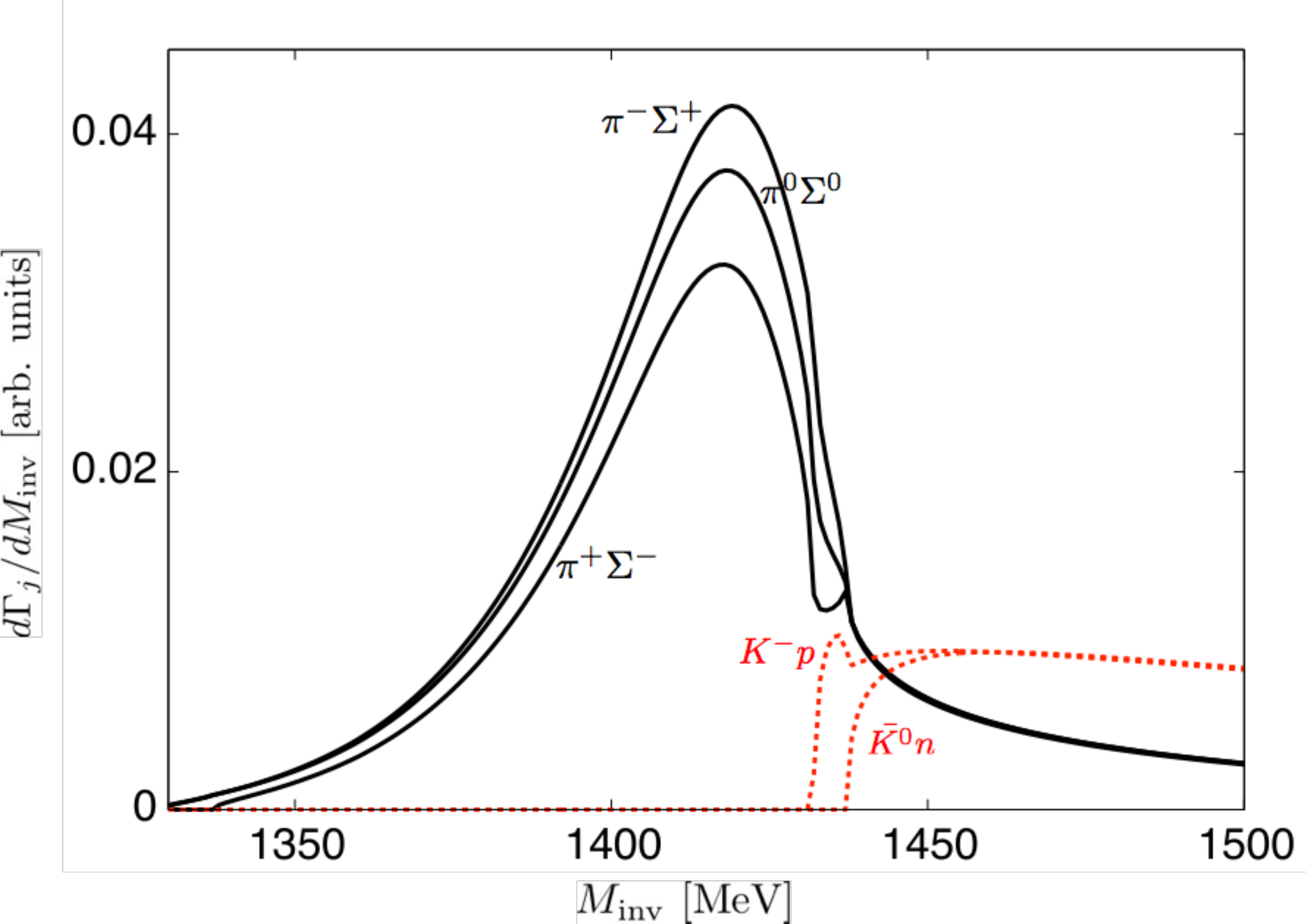}
\caption{(Color online) Invariant mass distribution of the decay $\Lambda_c^+\to \pi^+ MB$ near the $\bar{K}N$ threshold.  The solid line represents the spectrum for $\pi\Sigma$ channels and the dashed line represents the spectrum for $\bar{K}N$ channels. The meson-baryon amplitude is taken from Ref.~\cite{Ikeda:2012au}. }
\label{fig:massdis_IHW}  
\end{center}
\end{figure}

We show the spectra of three $\pi\Sigma$ channels in Fig.~\ref{fig:massdis_IHW}. From this figure, we find the $\Lambda(1405)$ peak structure around 1420 MeV. It is considered that the peak mainly reflects the pole at $1424-26i$ MeV. Because the initial state contains the $\bar{K}N$ channel with vanishing $\pi\Sigma$ component as shown in Eq.~\eqref{eq:hadronstate}, the present reaction puts more weight on the higher energy pole which has the strong coupling to the $\bar{K}N$ channel.

To proceed further, let us recall the isospin decomposition of the $\pi\Sigma$ channels~\cite{Nacher:1998mi}. The particle basis and the isospin basis are related as follows:
\begin{align}
|\pi^0\Sigma^0\rangle &= -\frac{1}{\sqrt{3}}|\pi\Sigma\rangle^{I=0}-\sqrt{\frac{2}{3}}|\pi\Sigma\rangle^{I=2}, \notag \\ 
|\pi^-\Sigma^+\rangle &= -\frac{1}{\sqrt{3}}|\pi\Sigma\rangle^{I=0}-\frac{1}{\sqrt{2}}|\pi\Sigma\rangle^{I=1}-\frac{1}{\sqrt{6}}|\pi\Sigma\rangle^{I=2}, \label{eq:piSig_rel} \\
|\pi^+\Sigma^-\rangle &= -\frac{1}{\sqrt{3}}|\pi\Sigma\rangle^{I=0}+\frac{1}{\sqrt{2}}|\pi\Sigma\rangle^{I=1}-\frac{1}{\sqrt{6}}|\pi\Sigma\rangle^{I=2}. \notag
\end{align}
In general reactions, the initial state of the $MB$ amplitude is a mixture of the $I=0$ and $I=1$ components.\footnote{In most cases, the small effect of $I=2$ can be neglected.}  The charged $\pi\Sigma$ spectra thus contains the $I=1$ contribution as well as the interference effects of different isospin components.

It is therefore remarkable that all the $\pi\Sigma$ channels have the same peak position in Fig.~\ref{fig:massdis_IHW}. This occurs because the present reaction picks up the $I=0$ initial state selectively, as explained in Sec.~\ref{sec:formulation}. In this case, the $I=1$ contamination is suppressed down to the isospin breaking correction, and hence all the charged states exhibit the almost same spectrum.\footnote{The small deviation is caused by the isospin violation effect in the meson-baryon loop functions.} The differences of the spectra, because of the $I=0$ filter in the present reaction, are much smaller than in photoproduction \cite{Moriya:2013eb,Moriya:2013hwg}, where the explicit contribution of the $I=0$ and $I=1$ channels makes the differences between the different $\pi\Sigma$ channels much larger, even changing the position of the peaks. In this respect, the $\Lambda_{c}\to\pi^{+}\pi\Sigma$ reaction is a useful process to extract information on the $\Lambda(1405)$, because even in the charged states (the $\pi^0 \Sigma^0$ automatically projects over $I=0$) one filters the $I=0$ contribution and the charged states are easier to detect in actual experiments.

The spectra for the $\bar{K}N$ channels are also shown in Fig.~\ref{fig:massdis_IHW}. In the $\bar{K}N$ channels, the peak of the $\Lambda(1405)$ cannot be seen, because the $\bar{K}N$ threshold is above the $\Lambda(1405)$ energy. However, the  enhancement near the threshold that we observe in Fig.~\ref{fig:massdis_IHW} is related to the tail of the $\Lambda(1405)$ peak. The shape of the $\bar{K}N$ spectrum, as well as its ratio to the $\pi\Sigma$ one, is the prediction of the meson-baryon interaction model. The detailed analysis of the near-threshold behavior of the $\bar{K}N$ spectra, together with the $\pi\Sigma$ spectra, will be important to understand the nature of the $\Lambda(1405)$.

\subsection{Spectrum above the $\bar{K}N$ threshold}

The spectrum above the $\bar{K}N$ threshold is also interesting. The LHCb Collaboration has found that a substantial amount of $\Lambda^{*}$s is produced in the $K^{-}p$ spectrum in the $\Lambda_{b}\to J/\psi K^{-}p$ decay~\cite{Aaij:2015tga}. Hence, the $K^{-}p$ spectrum in the weak decay process serves as a new opportunity to study the excited $\Lambda$ states.

For this purpose, here we adopt the model in Ref.~\cite{Oset:2001cn} for the meson-baryon final state interaction, which reproduces the $\Lambda(1670)$ as well as the $\Lambda(1405)$ in the $I(J^P)=0(1/2^-)$ channel. The pole position of the $\Lambda(1670)$ is obtained at $1678-20i$ MeV.\footnote{The present pole position is different from the one of the original paper \cite{Oset:2001cn}. This is because the original pole position is calculated with a physical basis though the present position is with isospin basis.}
Because the width of the $\Lambda(1670)$ is narrow, the pole of the $\Lambda(1670)$ also affects the invariant mass distribution of the $\Lambda_c^+$ decay. 

\begin{figure}[tb]
\begin{center}
\includegraphics[width=8cm,bb=0 0 655 463]{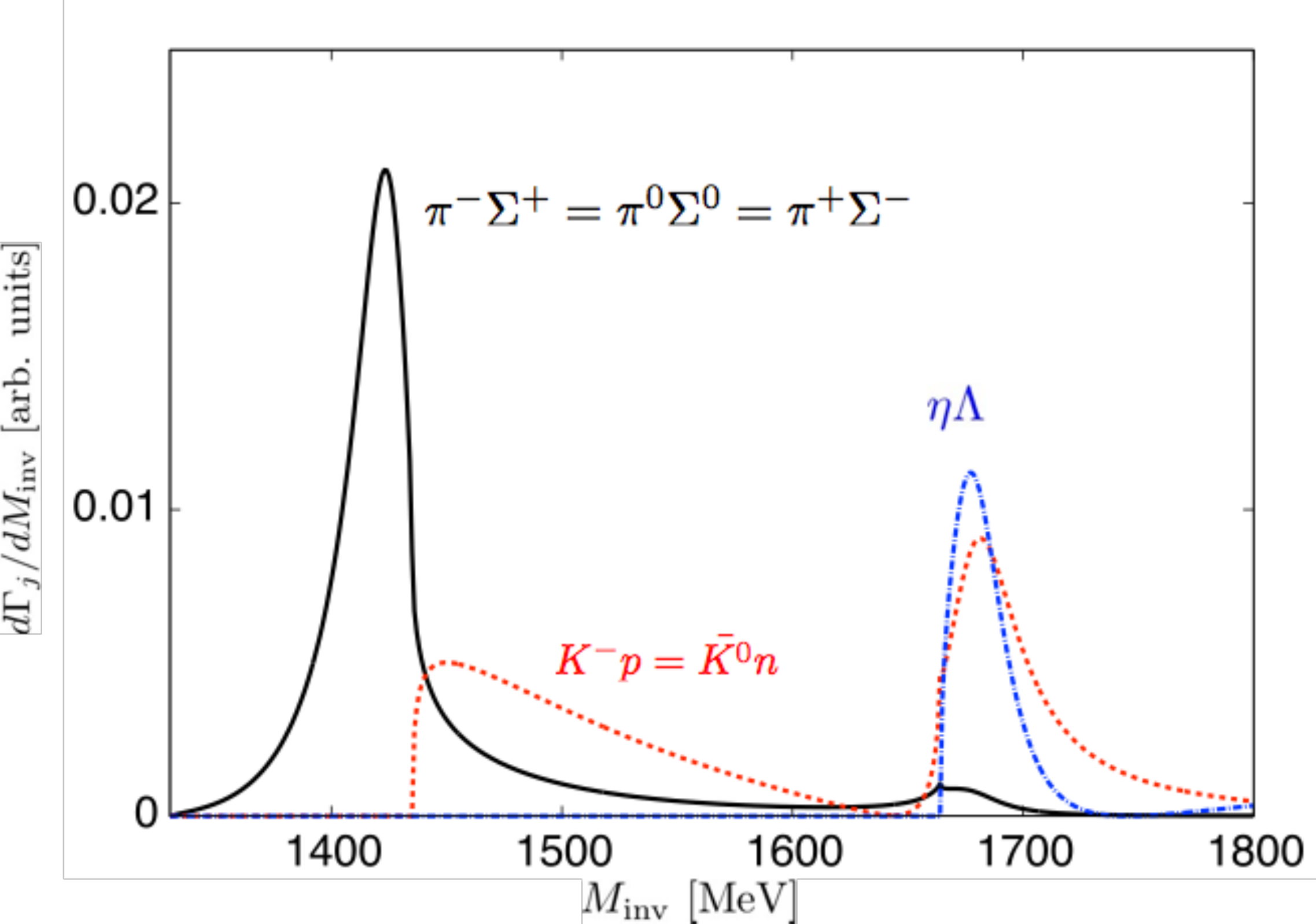}
\caption{(Color online) Invariant mass distribution of the decay $\Lambda_c^+\to \pi^+ MB$. The solid, dotted, and dash-dotted lines represent the $\bar{K}N=\{K^-p,\ \bar{K^0}n\}$, $\pi\Sigma=\{\pi^0\Sigma^0,\ \pi^-\Sigma^+,\ \pi^+\Sigma^-\}$, and $\eta\Lambda$ channels, respectively. The meson-baryon amplitude is taken from Ref.~\cite{Oset:2001cn}, where the $\Lambda(1520)$ contribution in $d$ wave is not included.}
\label{fig:massdis_ORB}  
\end{center}
\end{figure}

In Fig.~\ref{fig:massdis_ORB}, we show the invariant mass distribution of the $\Lambda_c^+$ decay into the $\pi\Sigma$, $\bar{K}N$, and $\eta\Lambda$ channels. Because the meson-baryon amplitude in Ref.~\cite{Oset:2001cn} does not include the isospin breaking effect, all the isospin multiplets $\{K^{-}p,\bar{K}^{0}n\}$, $\{\pi^{0}\Sigma^{0},\pi^{+}\Sigma^{-},\pi^{-}\Sigma^{+}\}$ provide an identical spectrum. Because the $\Lambda(1520)$ resonance in $d$ wave is not included in the amplitude, such contribution should be subtracted to compare with the actual spectrum.

As in the previous section, we find the $\Lambda(1405)$ peak structure in the $\pi\Sigma$ channel and the threshold enhancement in the $\bar{K}N$ channel. Furthermore, in the higher-energy region, we find the additional peak structure of the $\Lambda(1670)$ around 1700 MeV in all channels. Especially, the peak is clearly seen in the $\bar{K}N$ and $\eta\Lambda$ channels, as a consequence of the stronger coupling of the $\Lambda(1670)$ to these channels than to the $\pi\Sigma$ channel~\cite{Oset:2001cn}. The $\eta\Lambda$ channel is selective to $I=0$, and the $\Lambda(1520)$ production is kinematically forbidden. 

We expect that the structure of the $\Lambda(1670)$ can be analyzed from the measurements of the $\Lambda_c^+$ decay to the $\bar{K}N$ and $\eta\Lambda$ channels. 

\begin{figure}[tb]
\begin{center}
\includegraphics[width=8cm,bb=0 0 655 463]{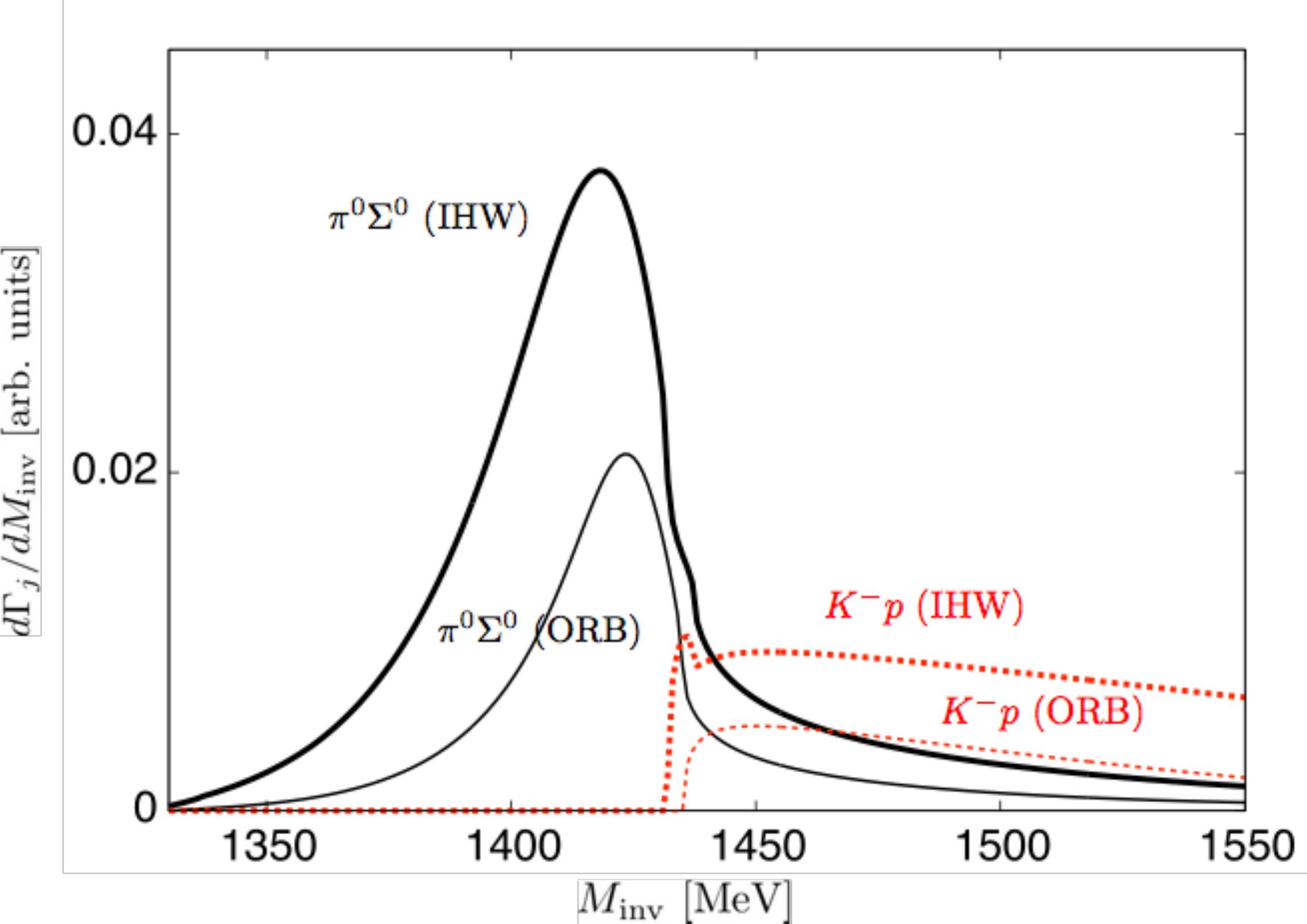}
\caption{(Color online) Comparison of the final-state interaction models of IHW~\cite{Ikeda:2012au} (thick lines) and ORB~\cite{Oset:2001cn} (thin lines) with a common normalization of the weak decay vertex $V_{P}$. The solid and dotted lines represent the $\pi^0\Sigma^0$ and $K^-p$ channels, respectively.}
\label{fig:massdis_IHW_ORB}  
\end{center}
\end{figure}

Finally, we examine the sensitivity of the spectrum to the $MB$ final-state interaction by comparing the spectra of Ref.~\cite{Ikeda:2012au} (hereafter called IHW) with that of Ref.~\cite{Oset:2001cn} (called ORB). In Fig.~\ref{fig:massdis_IHW_ORB}, we show the $\pi^{0}\Sigma^{0}$ and $K^{-}p$ spectra with the final-state interaction models of IHW and ORB with a common normalization of the weak decay vertex $V_{P}$. We find that the spectra of both channels in the IHW model are larger than those in the ORB model. This is caused by the difference of the loop function $G_{i}$ assigned before the final-state interaction in Eqs.~\eqref{eq:amplitude_isobasis_piSig} and \eqref{eq:amplitude_isobasis_KbarN}, which depend on the subtraction constants in each model.

\begin{figure}[tb]
\begin{center}
\includegraphics[width=8cm,bb=0 0 655 463]{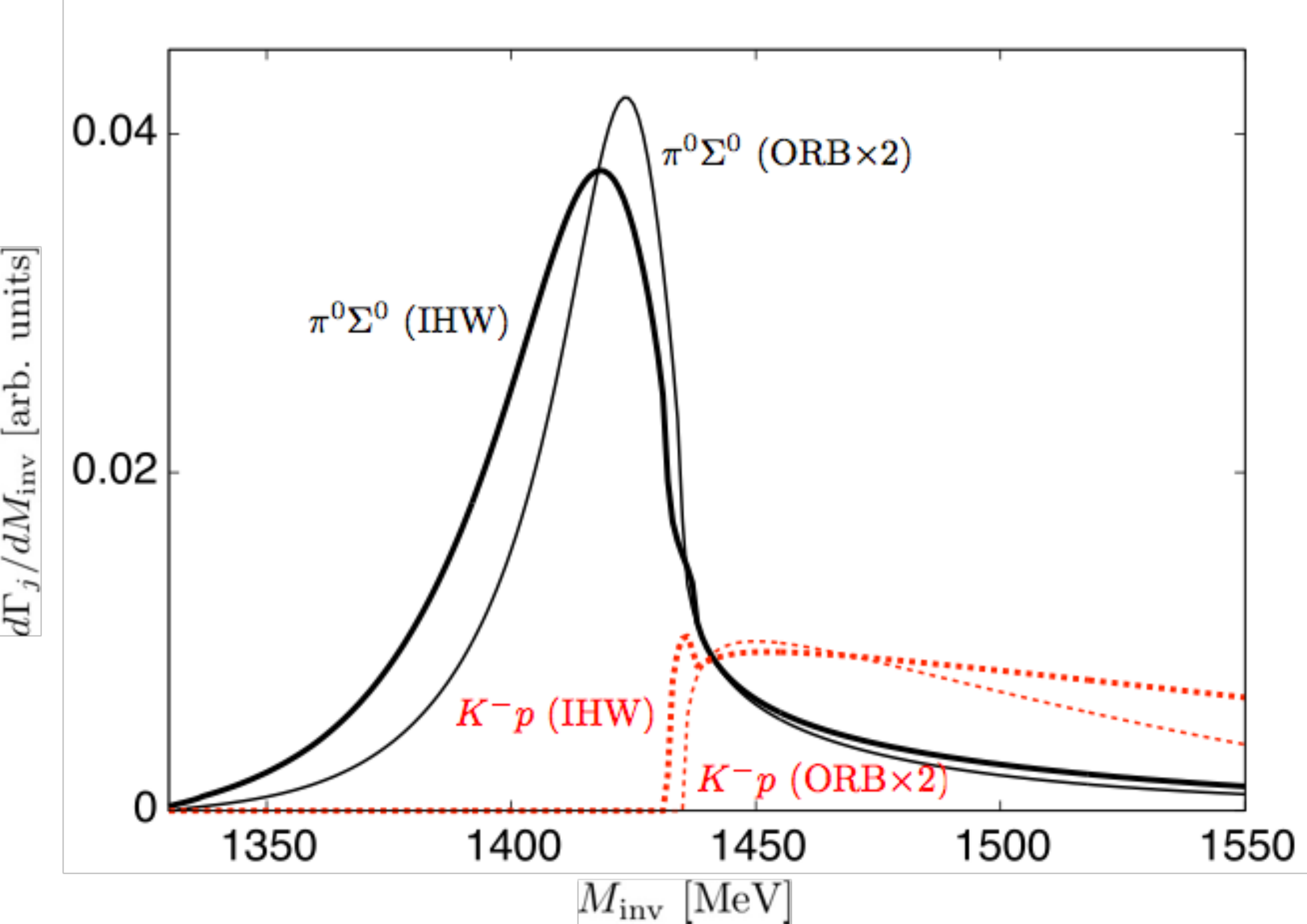}
\caption{(Color online) Comparison of the final-state interaction models of IHW~\cite{Ikeda:2012au} (thick lines) and ORB~\cite{Oset:2001cn} multiplied by factor two (thin lines). The solid and dotted lines represent the $\pi^0\Sigma^0$ and $K^-p$ channels, respectively.}
\label{fig:massdis_IHW_ORB2}  
\end{center}
\end{figure}

Because the absolute normalization of $V_{p}$ is not calculated here, we may renormalize the difference of the magnitude to compare the shape of the spectra. In Fig.~\ref{fig:massdis_IHW_ORB2}, we compare the same IHW spectra with the rescaled ORB spectra multiplied by factor two. We find that the peak position of the $\pi^{0}\Sigma^{0}$ spectrum of ORB is slightly closer to the $\bar{K}N$ threshold than that of IHW, and the decrease of the $K^{-}p$ spectrum of ORB above the threshold is steeper than that of IHW. This can be understood by the position of the higher-energy pole of $\Lambda(1405)$ in ORB at $1427-17i$ MeV, in comparison with $1424-26i$ MeV in IHW. This tendency is also seen in the comparison of two models in the $\Lambda_{b}\to J/\psi MB$ decay~\cite{Roca:2015tea}. This indicates that the $MB$ spectra in the weak decays well reflect the details of the meson-baryon scattering amplitude.

\subsection{Branching fractions}

Experimentally, the decay branching fractions of $\Lambda_{c}\to \pi^{+}MB$ are determined as~\cite{Agashe:2014kda}
\begin{align}
\Gamma(\Lambda_{c}\to pK^{-}\pi^{+},\text{nonresonant}) &= 2.8\pm 0.8, \% 
\label{eq:pKpi}\\
\Gamma(\Lambda_{c}\to \Sigma^{+}\pi^{+}\pi^{-}) &= 3.6\pm 1.0, \% \\
\Gamma(\Lambda_{c}\to \Sigma^{-}\pi^{+}\pi^{+}) &= 1.7\pm 0.5, \% \\
\Gamma(\Lambda_{c}\to \Sigma^{0}\pi^{+}\pi^{0}) &= 1.8\pm 0.8, \%
\end{align}
where the nonresonant component is obtained by subtracting the contributions from the $K^{*}(892)^{0}$, $\Delta(1232)^{++}$, and $\Lambda(1520)$ in the $K^{-}\pi^{+}$, $p\pi^{+}$, and $pK^{-}$ spectra, respectively. Taking the ratios of the central values, we obtain
\begin{align}
\frac{\Gamma(\Lambda_{c}\to \Sigma^{+}\pi^{+}\pi^{-})}
{\Gamma(\Lambda_{c}\to pK^{-}\pi^{+},\text{nonresonant}) } &= 1.3, \\
\frac{\Gamma(\Lambda_{c}\to \Sigma^{-}\pi^{+}\pi^{+})}
{\Gamma(\Lambda_{c}\to pK^{-}\pi^{+},\text{nonresonant}) } &= 0.6, \\
\frac{\Gamma(\Lambda_{c}\to \Sigma^{0}\pi^{+}\pi^{0})}
{\Gamma(\Lambda_{c}\to pK^{-}\pi^{+},\text{nonresonant}) } &= 
0.6.
\end{align}

In principle, these ratios can be calculated in the present model by integrating Eq.~\eqref{eq:massdis} over $M_{\rm inv}$. However, in the present calculation, we consider the process which is dominant in the small $M_{\rm inv}$ region, as explained in Sec.~\ref{sec:formulation}. In the large $M_{\rm inv}$ region, the processes other than Fig.~\ref{fig:Lambdac_decay} can contribute. Also, higher excited $\Lambda^{*}$~\cite{Feijoo:2015cca} and resonances in the $\pi^{+} M$ and $\pi^{+} B$ channels may play an important role.\footnote{The largest contributions from $K^{*}$, $\Delta$, and $\Lambda(1520)$ are subtracted in the data of Eq.~\eqref{eq:pKpi}.} In this way, the validity of the present framework is not necessarily guaranteed for the large $M_{\rm inv}$ region. 

Nevertheless, it is worth estimating the branching ratios by simply extrapolating the present model. The theoretical estimate of the ratio of the decay fraction is obtained as
\begin{align}
\frac{\Gamma_{\pi^-\Sigma^+}}{\Gamma_{K^-p}} = 
\begin{cases}
1.05 & (\text{Ref.~\cite{Ikeda:2012au}}), \\
0.95 & (\text{Ref.~\cite{Oset:2001cn}}). 
\end{cases} 
\label{eq:decayratio}
\end{align}
Given the uncertainties in the experimental values and the caveats in the extrapolation discussed above, it is fair to say that the gross feature of the decay is captured by the present model. We note that the difference of the charged $\pi\Sigma$ states in our model is of the order of the isospin breaking correction. The large deviation in the experimental data, albeit non-negligible uncertainties, may indicate the existence of the mechanisms which is not included in the present framework. It is worth noting that in the theoretical model of Ref.~\cite{Ikeda:2012au} the $\pi^- \Sigma^+ \pi^+$ channel has the largest strength as in the experiment.

Let us also mention the measured value of the branching fraction of $\Gamma(\Lambda_{c}\to \Lambda\pi^{+}\pi^{0}) = 3.6\pm 1.3 \%$~\cite{Agashe:2014kda}. Because $\pi^{0}\Lambda$ is purely in $I=1$, the present model does not provide this decay mode. The finite fraction of this mode indicates the existence of mechanisms other than the present process. In other words, the validity of the present mechanism for the $I=0$ filter can be tested by measuring the $\pi^{0}\Lambda$ spectrum in the small $M_{\rm inv}$ region. We predict that the amount of the $\pi^{0}\Lambda$ mode should be smaller than the $\pi\Sigma$ mode, as long as the small $M_{\rm inv}$ region is concerned.

\section{Summary}  \label{sec:summary}  

We have studied the $\Lambda_c\to \pi^+$ and $MB$ decay process, which is favored for several reasons, the CKM matrix, the color factor, the $ud$-diquark correlation, the number of $\bar{q}q$ creation, and the kinematical condition. Comparing with the similar decay $\Lambda_b\to J/\psi MB$, the $\Lambda_c$ decay benefits from the CKM-matrix element $|V_{ud}|\gg |V_{cd}|$ and the extra color factor in the $Wu\bar{d}$ vertex. The reaction has the virtue of filtering $I=0$, such that in the $\bar{K}N$ and $\pi\Sigma$ final states we do not have contamination of $I=1$ channels, which is not the case in other reactions, making the interpretation more difficult, and we can study better the properties of the $\Lambda(1405)$.

We have analyzed the $MB$ spectrum, taking into account the $MB$ final-state interaction by the chiral unitary approach. Near the $\bar{K}N$ threshold energy, the peak of the $\pi\Sigma$ spectra appears around 1420 MeV, reflecting the higher mass pole of the two poles of the $\Lambda(1405)$. Thanks to the suppression of the $I=1$ contamination, all three charge combinations of the $\pi\Sigma$ channels show almost the same spectrum. The model for the interaction of $\bar{K}N$ with $\pi\Sigma$ and other coupled channels allows us to make predictions for $\bar{K}N$ production and relate it to $\pi\Sigma$ production. 

Above the $\bar{K}N$ threshold, the peak of the $\Lambda(1670)$ can be observed in the $\bar{K}N$ and $\eta\Lambda$ channels. With the Bonn model of Ref.~\cite{Mai:2014xna}, the mass distribution did not show the $\Lambda(1670)$ peak in the $\Lambda_b\to J/\psi K^-p$ reaction. It thus becomes interesting to perform the $\Lambda_c\to \pi^+K^-p$ reaction, measuring the $K^-p$ mass distribution and see what one observes for this resonance, which is catalogued with four stars in the PDG \cite{Agashe:2014kda}.

In this way, we conclude that the $\Lambda_c\to \pi^+MB$ decay is an ideal process by which to study the $\Lambda$ resonances. We obtain the above-mentioned nice features based on the dominance of the mechanism shown in Fig.~\ref{fig:Lambdac_decay} for the $\bar{K}N$ and $\pi\Sigma$ channels. One should recall that, for the $\pi \Sigma$ spectra, integrated rates are available experimentally, but the peak of the $\Lambda(1405)$ has not been reported. The work done here should stimulate steps in this direction and also to measuring distributions that show the resonances discussed in the present paper.

\section*{Acknowledgments}

The authors thank the Yukawa Institute for Theoretical Physics, Kyoto University, where this work was initiated during the YITP-T-14-03 on ``Hadrons and Hadron Interactions in QCD.'' This work is supported in part by Open Partnership Joint Projects of JSPS Bilateral Joint Research Projects, JSPS KAKENHI Grant No. 24740152, the Yukawa International Program for Quark-Hadron Sciences (YIPQS), the Spanish Ministerio de Economia y Competitividad, European FEDER funds under Contracst No. FIS2011-28853-C02-01 and No. FIS2011-28853-C02-02, and the Generalitat Valenciana in the program Prometeo II, 2014/068. We acknowledge the support of the European Community-Research Infrastructure Integrating Activity Study of Strongly Interacting Matter (acronym HadronPhysics3, Grant Agreement No. 283286) under the Seventh Framework Programme of the EU.


%

\end{document}